# Memristor Threshold Logic: An Overview to Challenges and Applications

Alex Pappachen James
Senior Member, IEEE
Bioinspired Microelectronics Systems Lab
Nazarbayev University, Astana
Email: www.biomicrosystems.info/alex

*Abstract*—Once referred to as the missing circuit component, memristor has come long way across to be recognized and taken as important to future circuit designs. The memristor due to its ability to memorize the state, switch between different resistance level, smaller size and low leakage currents makes it useful for a wide range of intelligent memory and computing applications. This overview paper highlights broadly provides the uses of memristor in the implementation of cognitive cells for different imaging and pattern matching applications.

## I. Introduction

Memristor is one of the basic electronic device that completes the relationship between the basic parameter of flux and charge [1]. The capability of remembering or storing data in terms of resistance values is the main characteristic feature of memristors. This feature contributes to the development of non-volatile memories [2]. A summary of memristor devices and its applications is provided in the survey papers [3]–[5]. The use of threshold logic with memristors can help develop newer designs that integrate the concept of memories with that of computing [3]. The aim of the paper is to provide a targeted bibliographic overview on memristor circuits with a focus on threshold logic systems.

## II. Threshold Logic Memristors in Computing and Pattern Recognition

Memristors can switch between its resistive states by application of different levels of voltage across it. Since switches forms the basic idea of digital gates, they find application in development of logic gates. Further the ability to retain a resistance state even when the power is off enables the use of memristors as memories.

The behavior of the memristors resemble the principle of firing of neurons, and give indications of early learning mechanisms. The ability to update the weights and the resemblance with synaptic potentiation makes it a good candidate to be considered for mimicking biological neural circuits.

The generalization blocks for the logic gates and computing units is a topic that is explored with memory devices. The ability of memristors to accomodate so many different states or configurations makes it a useful candidate in learning based systems. The idea of a cognitive cell [6] that can be programmed to different logical representations including in-corporating the learning has been a basis to develop threshold logic systems as shown in Fig. 1.

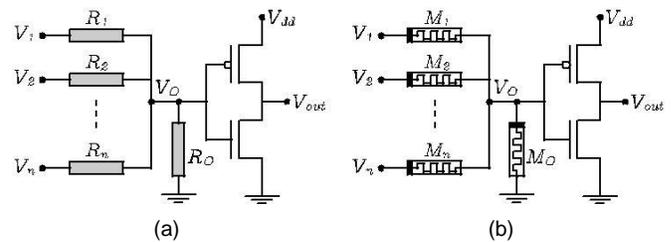

Fig. 1. The (a) cognitive cell proposed in [6] and its (b) memristor based counterpart.

### A. Character recognition

In [6], a hardware based memory cell is proposed for solving cognitive tasks. The proposed memory cell based architecture proves to be capable of avoiding the crossover wirings required in a neural network and it achieves the same functionality as of a neural network. Unlike other hardware based learning systems, the cell proposed in [6] does not require several iterations in order to learn a pattern but uses a bi-state weight model for quick learning. The learning potential of that architecture can be further improved if the bi-state model is extended to an n-state weight model using memristors. This may be achieved at the cost of additional driving/control circuits for the memristors. There are several approaches to incorporate and make use of learning in memris-tors for the application of character recognition some of these are [7] and [8].

### B. Logic Gates

The logic gates are the fundamental circuit cells required to build logic and arithmetic operators in modern processors. The conventional CMOS logic cells designs becomes challenging in sub 20nm technology sizes. Memristors due to its properties of low area on chip, low leakage currents and ability to emulate neural firing mechanisms make it a useful alternative to conventional CMOS gates. Some core designs include that based on switching logic [9], ratioed logic [10], state logic [11], and threshold logic [12], [13].

## C. Image Processing

In [14], the idea of cognitive cells were utilized for locating lesion probable regions in film mammography. Here, the cognitive cell parameters were adjusted according to the local and global input image statistic. In [15], [16], novel cognitive cell based architectures are used to find edges in digital/analog images. It was observed that the edge response obtained (Figure. 2(a)) by the use of cognitive cells was satisfactory. Exploiting the advantage of high processing speeds offered by cognitive cell based learning, the work in [17] proposed a method for real-time processing of medical data. The input considered were intraoperative MRI images (Figure. 2(c)) and they were processed to generate the heart activity graph (Figure. 2(d)).

## D. Object Recognition

Later in [18], the architecture proposed in [6] was studied for memristors. They studied memristor based architectures for implementing digital logic gates. Having the basic gates implemented using memristors, we can develop systems that perform higher level computing at great speeds. The work illustrated in [19] is an example proving that memristor based systems can be deployed for performing complex cognitive tasks like object detection and tracking shown in Figure.2(b).

## E. Signal Processing

The work in [20], proposes an implementation of Fast Fourier Transform and Vedic Multiplication using memristors. They highlighted the importance of memristors by reporting the lower chip area, THD and controllable leakage power.

## F. Speech Recognition

The speech recognition represents one of the applications where the real-time processing of features becomes very important. The hardware processing of signals through memristor pattern recognisers can increase the overall performance of speed. Some of recent examples of attempts to build real-time speech recognition include that based on threshold logic [21] and neural circuits [22].

## G. Face recognition

The recognition of the faces require the extraction of facial features from the images and would need to be robust to changes in natural variability. This problem represent a class of pattern recognition problem that is often limited by inability to acquire sufficient number of training samples. In such cases, a hardware based approach of processing patterns and ensuring a high speed match would enable on-chip face recognition methods. Some of the recent approaches include the development of HTM spatial pooler [23], threshold logic based pattern matching [24] and neuromorphic approach to face recognition [25].

## III. MEMRISTOR MODELS

There exists a wide range of memristor models in use today for SPICE, Verilog, MATLAB and C++. While, there is attempts to generalize the models, in practice it needs to be carefully used based on the specific type of memristor used, and the circuit functionality that the designer is aiming towards. The paper [3] provides some example models that often provides the most agreed characteristics of memristors. Among the recent works, two of the models that takes into effect time related aspects [26] and that attempts to improve the processing speed [27] needs attention. In particular, for circuit designers the ability to run large scale circuits would become an increasingly important issue, and going further to practical realization the accuracy of these models will require further examination.

The early form of the memristor model for getting started on SPICE model is the HP memristor model:

```
************************************************   *
HP Memristor SPICE Model
For Transient Analysis only
created by Zdenek and Dalibor Biolek
************************************************   *

Ron, Roff - Resistance in ON / OFF States
*  Rinit - Resistance at T=0
*  D - Width of the thin film
*  uv - Migration coefficient
*  p - Parameter of the WINDOW-function for
modeling nonlinear boundary conditions
*  x - W/D Ratio, W is the actual width
of the doped area (from 0 to D)
************************************************
.SUBCKT memristor plus minus PARAMS: + Ron=100
Roff=16K Rinit=11K D=10N uv=10F p=10
************************************************
*   DIFFERENTIAL EQUATION MODELING    *
************************************************
Gx 0 x value=I(Emem)*uv*Ron/D**2*f(V(x),p)
Cx x 0 1 IC=(Roff-Rinit)/(Roff-Ron)
Raux x 0 1T

    ************************************************
*   RESISTIVE PORT OF THE MEMRISTOR   *
************************************************
Emem plus aux value=-I(Emem)*V(x)*(Roff-
Ron) Roff aux minus Roff
************************************************   *
FLUX COMPUTATION *
************************************************
Eflux flux 0 value=SDT(V(plus,minus))
************************************************   *
CHARGE COMPUTATION *
************************************************
```

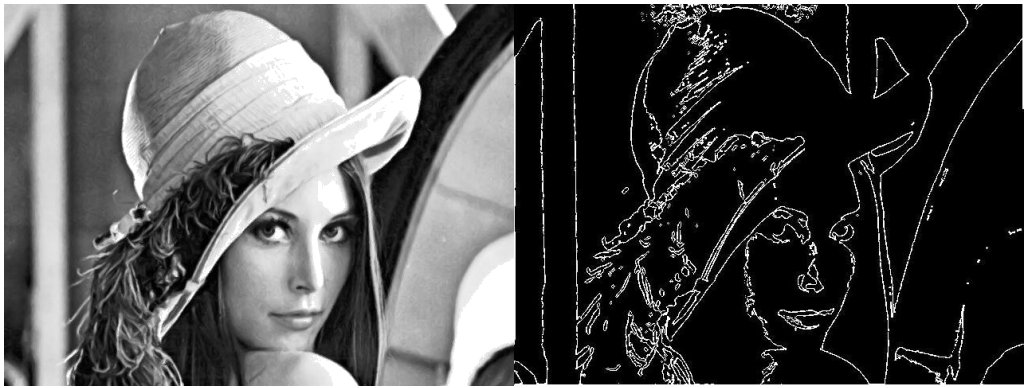

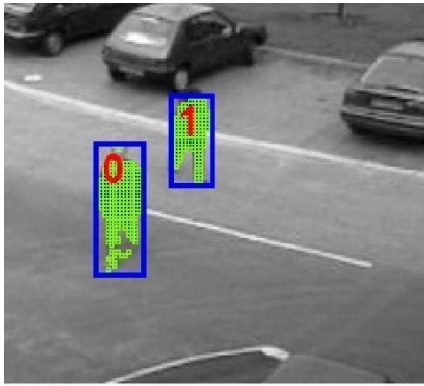
(b) Object Detection and Tracking [19]

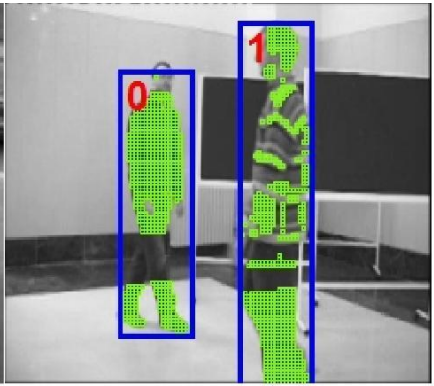

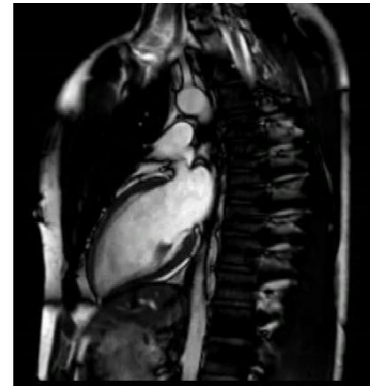
(c) iMRI data [17]

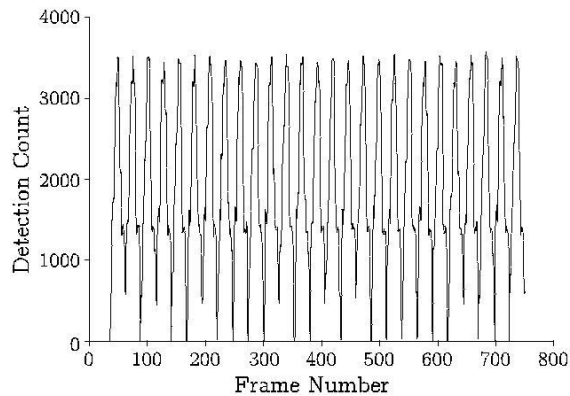
(d) Heart rate measured from iMRI [17]

Fig. 2. Example results obtained for cognitive cell based systems.

```
Echarge charge 0 value=SDT(I(Emem))
************************************************
WINDOW FUNCTIONS
FOR NONLINEAR DRIFT MODELING *
************************************************
window function, according to Joglekar
.func f(x,p)=1-(2*x-1)**(2*p)
proposed window function
;.func f(x,i,p)=1-(x-sttp(-i))**(2*p)
.ENDS memristor
```

In this model, it can be seen that resistor states requires to be defined. While its assumed that they remain constant, in real devices this does have a variability to it. Further, depending on the modification one makes to the device structure and material stack, the model itself would require further modifications to account for variability in the device characteristics.

## IV. DISCUSSION AND CONCLUSIONS

The device level explorations remain as one of the the most active areas of research. There have been a large number of devices from as early as 1970's that provide indications of memristive behaviors. There has been also debates on what can be classified as memristors to distinguish itself from purely a resistive switching behavior. The reliability of devices to be used along with conventional CMOS technologies is also a topic of importance, in particular the integration to commercially available technologies remain challenging for small devices sizes.

The array of memristors in crossbar architecture has proven to be one of the ways ahead to scale the memristor networks. Since the arrays can be stacked across 3 dimensions in rows and columns, the resistance changes can be readout at high speeds. While, the resistive nature of the network prevents a truly parallel processing due to the loading effects and leak path problems. In addition to these, the circuit level challenges include the development of interface circuits and programming circuits that enable the one-chip read-write operations of the memristors in array. In a large scale array these are not trivial issues, and require a range of techniques to be explored.

While the use of memristor circuits for a wide range of application is promising there are also quite a wide range of challenges. Many of these challenges are application specific and they vary based on the scale of the problem attempted. The circuit solutions required for a logic circuit design is different to that from a neural circuit emulation. Further, the higher levels of integration between the analog and digital signals at a varied voltage and power levels make a reliable circuit development challenging.

There has also been several advances in the simulation models for the memristors. However, many of the models available today needs to be tuned for the specific type of memristors used to be close to a realistic device. The variability between the memristive devices is still on a higher end than the more stable CMOS devices, and this requires further development of device models. The simulations with the existing memristor device models is challenging for large arrays and in many cases not possible in tools based on SPICE for system level simulations. While there are these ongoing challenges, the field is growing and many of these challenges pose as opportunities to build new technologies and applications.

This paper provided an overview of the memristor circuits in relation to threshold logic circuits. The increased interest in integrating memory with computing makes memristors a promising device for future computing platforms. The pos-sibility to integrate and mimic learning algorithms makes the memristor circuits versatile to build artificial intelligent systems.